\documentstyle[aps,prl,epsfig,multicol]{revtex}
 \draft
\begin{document}
\title{Elastic moduli of vortex lattices within nonlocal London model\\
 }
\author{P. Miranovi\' c$^1$ and V. G. Kogan$^2$}
\address{$^1$ Department of Physics, University of Montenegro,
P.O. Box 211, Yugoslavia\\
$^2$ Ames Laboratory and Department of Physics, Ames 50011}
\date{\today}
\maketitle
\begin{abstract}
Vortex lattice (VL) elastic response is analyzed within nonlocal London model.
 The squash modulus turns zero at the field $H_{\Box}$ where VL undergoes
square-to-rhombus structural transition.  In the field region $H>H_{\Box}$
where the square  VL is stable, the rotation modulus turns zero at a field
$H_r$ indicating instability of the square VL with respect to rotations. The
shear modulus depends on the shear direction; the dependence is strong in
the vicinity of $H_{\Box}$ where the square VL is soft with respect to the
shear along [110].  The  $H$ dependences of elastic  moduli are evaluated
  for LuNi$_2$B$_2$C. 
 \end{abstract}

 \begin{multicols}{2}

  The theory of elasticity for vortex lattices (VL) in type-II
superconductors is commonly constructed in a manner similar to the
standard elasticity \cite{Landau}. One introduces the local VL displacement
${\bf u}$ from equilibrium and expands the elastic energy $E$ in a series
of derivatives $u_{i,k}\equiv \partial u_i/\partial x_k$.
 The similarity, however, ends already at this stage. Elastic energy of an
element of a solid does not depend on rigid rotations of this element, so
that $E$   depends only on symmetric combinations
$u_{ik}=(u_{i,k}+u_{k,i})/2$   which form the strain
tensor, whereas the antisymmetric combinations
$\omega_{ik}=(u_{i,k}-u_{k,i})/2$  representing rotations are irrelevant.
This property can be tracked back to the basic isotropy of the space in
which  solids are situated. 

Vortices, however, are hosted by crystals which at best have the cubic
symmetry. In other words, the ``space in which  vortices live" is never
isotropic. The crystal anisotropy affects  intervortex interactions 
in a trivial way through the tensor $m_{ik}$ of ``superconducting masses"
which determines the anisotropy of the London penetration depth $\lambda$
and the coherence length $\xi$. For all symmetries except cubic (for
which $m_{ik}$ reduces to   $\delta_{ik}$), rigid rotations of
 VL cause an energy change even for rotations about the
direction of vortex axes (that of the magnetic induction 
${\bf B}$) \cite{kogan_prl_1989}. 

Moreover, the current and field distributions around the vortex core are
affected by the underlying crystal in a more subtle manner than via the 
$\lambda$ anisotropy. In full, this influence can be described only by
the microscopic theory. In materials with $\lambda /\xi=\kappa \gg 1$ the
problem is simplified at distances $r\gg\xi$  relevant for
intervortex interactions in fields well below the upper critical field
$H_{c2}$. This is done in the frame of the London theory corrected for the
nonlocality of the relation between the current and the vector potential.
The kernel of this nonlocal relation provides the formal bridge between the
Fermi system of   electrons in a given crystal and interacting vortices in
the superconducting condensate \cite{kogan_prb_1996,non}. In general, this
kernel depends on the Fermi surface and on the symmetry of the order
parameter \cite{Franz}; for simplicity,  we consider here 
the isotropic order parameter, so that all superconducting anisotropies are
determined by the Fermi surface.  We focus on the simplest part of the VL 
elastic response and consider only straight and parallel
vortices thus restricting ourselves to the two-dimensional (2D) problems. 
In fact, we calculate only ``shear" moduli (meaning deformations
preserving the vortex density, div${\bf u}\equiv u_{i,i}=0$). These moduli
are practically non-dispersive, that allows us to consider uniform
deformations (${\bf k}=0$) \cite{Brandt_nonloc}.

We begin  with the general discussion of the deformed VL in cubic or
tetragonal crystal in   fields along the crystal direction $c$, which
allows us to correctly count independent elastic constants. Then we evaluate
the moduli using the London model with nonlocal corrections for
LuNi$_2$B$_2$C. As expected, the squash modulus, corresponding to the
transformation of the rhombic unit cell to the square, turns zero at the
transition field $H_{\Box}$\cite{Baruch}. We find also novel features of the
elastic response. The shear moduli  depend strongly  on the displacement
direction, the shear ``polarization": the shear polarized along [110] is much
softer than others near $ H_{\Box}$.  The  rotation modulus of the square VL
turns zero at a field $H_r > H_{\Box}$, indicating instability of certain
orientations of the square VL. Then we discuss implications of our results
for the VL physics. \\ 
 
The general form of the elastic energy  in terms of strains and
rotations is  $ E=(\lambda_{ijkl}u_{ij}u_{kl}+
\eta_{ijkl}\omega_{ij}\omega_{kl})/2+\zeta_{ijkl}u_{ij}\omega_{kl}$ 
\cite{kogan_prl_1989}.  This form is not convenient for counting 
number of independent coefficients because one has to consider symmetry
restrictions upon three 4th rank tensors. A more
direct approach is to forgo the splitting of $u_{i,k}$ into symmetric and
antisymmetric parts (which no longer simplifies the problem) and to write 
the energy in terms of $u_{i,k}$: 
\begin{equation}
E= \gamma_{iklm}u_{i,k}u_{l,m}/2 \,\,. \label{general}
\end{equation}
 The tensor $\gamma_{iklm}$ is not  symmetric relative to
$i\leftrightarrow k$ and $l\leftrightarrow m$, but preserves the 
property of the standard elasticity tensor $\gamma_{ik,lm}=\gamma_{lm,ik}$.
The symmetries of this tensor and the number of its independent
components are determined both by the crystal and by the equilibrium VL
structure. In other words, $\gamma_{iklm}$ has only the symmetries 
common to the  crystal and the equilibrium VL.

The equilibrium VL in cubic or tetragonal crystals in fields above
$H_{\Box}$   along   [001] has a square unit cell
with sides along [110] and [${\bar
1}10$] \cite{yaron,eskild,wilde,yethi,paul,vinnikov,eskild1}. We choose the
coordinates $x,y$ in the $ab$ plane as [100]
and [010].  Then, $x$ may enter the indices of   $\gamma$ only even
number of times, and the components obtained from each other by 
replacements $x\leftrightarrow y$ are equal:
\begin{eqnarray}
E&=&{1\over 2} \,[\gamma_{xxxx}(u_{x,x}^2+
u_{y,y}^2)+2\gamma_{xxyy}u_{x,x} u_{y,y}\nonumber\\
&+&  \gamma_{xyxy}(u_{x,y}^2+ u_{y,x}^2) 
+ 2\gamma_{xyyx}u_{x,y}u_{y,x}]\,. \label{FF}
\end{eqnarray}
 After  excluding compressions ($B$ is constant, div${\bf u}=u_{x,x}+
u_{y,y}=0$) we are left with {\it three} constants:
\FL
\begin{eqnarray}
E&=& (\gamma_{xxxx}-\gamma_{xxyy}) u_{x,x}^2 
 +{1\over 2} \gamma_{xyxy}(u_{x,y}^2+ u_{y,x}^2)\,\nonumber\\ 
&+& \gamma_{xyyx}u_{x,y}u_{y,x}\nonumber\\
&=&{1\over 2}\,[\gamma_1u_{x,x}^2+\gamma_2(u_{x,y}^2+ u_{y,x}^2) 
+ 2\gamma_3 u_{x,y}u_{y,x}].
\end{eqnarray}
 
 How to classify the three constants, to a large extent, is a question of
semantics.  For our particular problem, a  uniform deformation defined as 
\begin{equation}
{\bf u} = \mu (x\,{\bf e}_x-y\,{\bf e}_y)\,,\label{squash}
\end{equation}
 ($\mu$ is a small constant and ${\bf e}_x,{\bf e}_y$ are unit vectors) is
of a special interest because it is this deformation which transforms the
square VL above $H_{\Box}$ into a rhombic one for $H < H_{\Box}$. Since this
is a second order phase transition, the corresponding modulus must vanish 
at $H=H_{\Box}$ \cite{Baruch}. This deformation was  named 
``squash"\cite{kogan_prl_1989}. For the displacement (\ref{squash}),
$u_{x,x}=-u_{y,y}= \mu$, whereas $ u_{y,x}=u_{x,y}=0$. Then, the part of the
energy related to the squash is $E_{sq}= \gamma_1 \mu^2/2$. In other words,
$\gamma_1=2(\gamma_{xxxx}-\gamma_{xxyy})$ can be called the squash modulus and
denoted as $C_{sq}$.   
 
Consider now a uniform shear polarized along  $x$, 
\begin{equation}
{\bf u} = \nu \,y\,{\bf e}_x\,, 
\label{shearX}
\end{equation}
for which only $u_{x,y}\ne 0$. The energy is $E_{x}=\gamma_2 u_{x,y}^2/2=
\gamma_2 \nu^2/2$. The energy 	is the same for the $y$ polarization:
$E_y=E_x=\gamma_2\nu^2/2$.  Hence, $\gamma_2$ is the shear modulus for either
$x$ or $y$ polarizations; below we complement  the traditional notation   with
polarization index: $\gamma_2=C_{66,x} =C_{66,y}$. 

Let now the shear be directed along some direction ${\bf e}_{x^{\prime}}$
 so that the displacement is given by  ${\bf u} = \nu \,y\,^{\prime}\,{\bf
e}_{x^{\prime}}$, where
\begin{equation}
x\,^{\prime}=x\cos\theta+y\sin\theta\,, \quad
y\,^{\prime}=-x\sin\theta+y\cos\theta\,
\end{equation}
and $\theta$ is the angle between $x\,^{\prime}$ and $x$.  
In the old coordinates,
\begin{equation}
{\bf u} = \nu \,(-x\sin\theta+y\cos\theta)\,({\bf e}_
x\cos\theta+{\bf e}_y\sin\theta)\,.
 \label{sh_gen}
\end{equation}
Then $u_{x,x}=-u_{y,y}=-\nu\sin \theta\cos
\theta$, $u_{x,y}=\nu\cos^2\theta$, $u_{y,x}=-\nu\sin^2\theta$, and
\begin{equation}
 E(\theta)=  {\nu^2\over
2}[\gamma_2+(\gamma_1-2\gamma_2-2\gamma_3)\sin^2\theta\, cos^2\theta]\, . 
\end{equation}
In particular, $E(\pi/4)=\nu^2(\gamma_1+2\gamma_2-2\gamma_3)/8$. Since all
$u_{i,j}^2=\nu^2/4$,  the factor $\gamma_1+2(\gamma_2-\gamma_3)$ is the
modulus for the shear polarized along [110].  
   Thus, in addition to  $C_{sq}=\gamma_1$, there are two independent 
shear moduli in the problem: $C_{66,x}=\gamma_2$ and
$C_{66}(\pi/4)=\gamma_1+2(\gamma_2-\gamma_3)$. 
 
  Consider now a rigid rotation 
\begin{equation}
{\bf u} = \omega (x\,{\bf e}_y-y\,{\bf e}_x)\,,\label{rot}
\end{equation}
for which  $u_{y,x}=- u_{x,y}=\omega$, whereas $u_{x,x}=0$.  Then, 
  $E_r = \omega^2(\gamma_2-\gamma_3)= C_r \omega^2/2$. 
 In other words, the rotational modulus is $C_r=2(\gamma_2-\gamma_3)$.
Thus, the three independent moduli can also be chosen as 
\begin{equation}
 C_{sq}=\gamma_1,\quad C_{66,x}=\gamma_2,\quad {\rm and}\quad
C_r=2(\gamma_2-\gamma_3).
\end{equation}
Note that 
\begin{equation}
C_{66}(\pi/4)=C_{sq}+C_r\,. \label{pi/4}
\end{equation}
  The choice of a particular set of elastic moduli is a matter of
convenience  in a problem at hand.

The above enumeration of the elastic constants pertains only for $H >
H_{\Box}$ (parallel to $c$) where both the crystal and the VL have the
square symmetry in the $xy$ plane. In fields under $H_{\Box}$, the
equilibrium VL is rhombic, and the replacement $x\leftrightarrow y$ is no
longer a symmetry operation. We then obtain instead of Eq. (\ref{FF}): 
\begin{eqnarray}
&E&={1\over 2} \,(\gamma_{xxxx}u_{x,x}^2+\gamma_{yyyy} u_{y,y}^2 +
2\gamma_{xxyy}u_{x,x} u_{y,y}\nonumber\\ 
 &+& \gamma_{xyxy}u_{x,y}^2
+\gamma_{yxyx} u_{y,x}^2)
+ \gamma_{xyyx} u_{x,y}u_{y,x}\,. \label{FFF}
\end{eqnarray}
For an incompressible VL, we obtain four independent constants.  We can
choose them  as the squash $C_{sq}$, two   shears
$C_{66,x} \ne C_{66,y}$, and the rotation   $C_r$. \\
 
We now turn to evaluation of the moduli using the London equations
corrected for nonlocality, which are obtained from the
  BCS theory for superconductors  with isotropic
gap\cite{kogan_prb_1996}:
\begin{equation}
{4\pi\over c}j_i=-{1\over \lambda^2}\,(m_{ij}^{-1}-\lambda^2
n_{ijlm}k_lk_m)a_j\,.
\label{Londnon}
\end{equation}
 Here, $m_{ij}  $ is the normalized mass tensor 
($\det m_{ij}=1$) and  $\lambda$ is the average penetration depth
($\lambda^3=\lambda_1\lambda_2\lambda_3$, $\lambda_1=\lambda\sqrt{m_1}$,
...)  Further,
${\bf a}={\bf A}+\phi_0\nabla\theta/2\pi$ and
$\phi_0$ is flux quantum. Tensor $n_{ijlm}$ is defined as
\begin{equation}
n_{ijlm}={3\hbar^2\langle v_iv_jv_lv_m\rangle
\over 4\langle v^2\rangle \Delta^2\lambda^2}\,\gamma (T,\tau)\,,\quad 
\gamma={\Delta^2\sum\beta^{-2}{\beta_1}^{-3}
\over\sum \beta^{-2}{\beta_1}^{-1}}, 
\end{equation}
where $\bf v$ is the Fermi velocity and the brackets
$\langle\ldots\rangle$ stand for averages over the Fermi surface;  
  $\beta^2=\Delta^2+\hbar^2\omega^2$, $\Delta(T)$ is the uniform BCS gap,
$\hbar\omega=\pi T(2n+1)$ with an integer $n$, and sums   run over
  $\omega>0$; $\beta_1=\beta+\hbar/2\tau$ with  $\tau$ being the 
scattering time due to nonmagnetic impurities. In dirty superconductors
 $\gamma \sim (\tau\Delta/\hbar)^2\rightarrow 0$, i.e., nonlocal 
effects vanish. In the clean limit $\gamma$ ranges from $2/3$ at $T=0$ to
$\approx 0.3$ at $T=T_c$. 

The flux quantization  $\nabla\times {\bf a}={\bf h}-
\phi_0\;{\bf e}_z\delta({\bf r})$ combined with Eq. (\ref{Londnon})
gives the field  of a single vortex along   $c$   of cubic or
tetragonal materials:
\begin{equation}
h_z({\bf k})={\phi_0\over
1+\lambda^2k^2+\lambda^4(n_{xxyy}k^4+dk_x^2k_y^2)}\,,
\label{field}
\end{equation}
where $d=2n_{xxxx} -6n_{xxyy}$. The free energy density is a sum of
pairwise interactions of vortices \cite{book}; in the reciprocal space we
have 
\begin{equation}
F={B^2\over 8\pi \phi_0}\sum\limits_{{\bf q}}s({\bf q})
h_z({\bf q})\,,
\label{free}
\end{equation}
 where ${\bf q}$ form the reciprocal VL.  The factor $s({\bf q})$ is
introduced to deal with the  shortcoming of the London model  which
disregards spatial variations of the order parameter in vortex cores.  At low
temperatures  of our interest, we use
$s(q)=\exp(-q^2\xi^2/2)$\cite{brandt_cut_1}. By and large, this form of the
cutoff is confirmed in   neutron scattering experiments from which the
form-factor $|s({\bf q})h_z({\bf q})|^2$ can be extracted\cite{Gammel}.  We
consider the typical  case of a platelet sample with the $c$ axis normal to
the flat face in  fields along $c$. Then  minimization of $F$  at a given
$B$ yields the equilibrium VL, whereas its elastic properties are evaluated by
considering deviations from the  minimum.   

 For high-$\kappa$ materials, away of the lower critical field,   the
interaction of vortices extends over several intervortex spacings. This gives
rise to the nonlocal elastic response of VL's \cite{Brandt_nonloc}, for
which the elastic moduli, in general, depend on the deformation length
scale (on the wave vector ${\bf k}$).  However, even in isotropic case the
shear modulus is practically ${\bf k}$ independent. We focus
here on the nondispersive modes, i.e., we consider the elastic response to
   deformations with ${\bf k}=0$. 
 
We begin with the equilibrium square VL with  basis
  vectors
 ${\bf a}_1=a_0 ( {\bf e}_x+ {\bf e}_y)/\sqrt{2}\,,\quad 
{\bf a}_2=a_0 
({\bf e}_y-{\bf e}_x)/\sqrt{2}.$ 
 The reciprocal lattice is 
\begin{equation}
q_x=\pi\sqrt{2}(m+n)/a_0\,,\qquad q_y=\pi\sqrt{2}(m-n)/a_0\,,
\end{equation}
with integers $m$ and $n$. 
 For  the 2D uniform deformations, the displacement is $u_i=u_{i,j}x_j$ with
constant $u_{i,j}$'s. Since the VL cell area is fixed at a given $B$,
$u_{x,x}+u_{y,y}=0$\cite{Landau}. The deformed cell is given by
$a_{\alpha i}^{\prime}= a_{\alpha i}+ u_{i,j}a_{\alpha j}$ with
$\alpha=1,2$. It is readily shown that the reciprocal VL  is  also 
homogeneously deformed
${\bf q}^\prime={\bf q}+{\bf u}^*$ with 
\begin{equation}
u^*_{  i}=-u_{j,i}q_{  j}\,   \label{u*}
\end{equation}
 (with the tensor of derivatives being transposed).
 
We now expand the  intervortex interaction $V({\bf q}^\prime)=
s(q^\prime)h_z({\bf q}^\prime)/\phi_0$ in powers of ${\bf u}^*$:   
\begin{equation}
V({\bf q}^\prime)=V({\bf q})+
u_i^*{\partial V\over \partial q_i}+\frac{1}{2}u_i^*u_j^*
\frac{\partial^2 V}{\partial q_i\partial q_j}+\ldots\,.
\end{equation}
The term linear in $u_i^*$ in the sum (\ref{free}) must vanish since
$u_i^*=0$ correspond to equilibrium.
 We then obtain for the elastic energy $  E=F\{{\bf
q^\prime}\}-F\{{\bf q}\}$:  
\begin{equation}
 E= {B^2\over 16\pi }\sum\limits_{{\bf q}}\Big(\frac{\partial^2
V}{\partial q_i\partial q_j}\,q_{\mu}q_{\nu}\Big)u_{\mu,i}u_{\nu,j}\,.
 \end{equation}
 
 For the squash deformation (\ref{squash})   this gives
 \begin{equation}
C_{sq}={B^2\over 8\pi}\sum\limits_{\bf q}
\left (q_x^2{\partial^2 V\over \partial q_x^2}+q_y^2{\partial^2 V\over
\partial q_y^2}- 2q_xq_y{\partial^2 V\over \partial q_x\partial q_y}
\right ). \label{C_{sq}}
\end{equation}
Similarly, we obtain for the shear (\ref{shearX}) and rotation
(\ref{rot}):
\begin{eqnarray}
&&C_{66,x}={B^2\over 8\pi}\sum\limits_{\bf q}
q_x^2\,{\partial^2 V\over \partial q_y^2}\,,
 \label{C_{66,x}}\\
&&C_{r}={B^2\over 8\pi}\sum\limits_{\bf q}
\left(
q_y^2{\partial^2 V\over \partial q_x^2}
+q_x^2{\partial^2 V\over \partial q_y^2}-
2q_xq_y{
\partial^2 V\over\partial q_x \partial q_y}\right).\label{C_r}
\end{eqnarray}
 These three moduli form the complete set of 2D independent elastic
constants for the square VL in cubic or tetragonal crystals in fields
$H > H_{\Box}$ along [001]. If needed, $C_{66}(\pi/4)$ can be evaluated
using Eq. (\ref{pi/4}).

In a similar manner one can obtain the four moduli for the rhombic  VL 
  below $H_{\Box}$. The moduli can be calculated numerically by
evaluating the   sums in Eqs. (\ref{C_{sq}} - \ref{C_r}). The
results of such a calculation for  LuNi$_2$B$_2$C (with parameters given in
Ref. \onlinecite{non}) are shown in Fig.1, where the field
dependence of the moduli is shown. It is worth  noting that  $C_r$ turns zero
at some field $H_r$ which is about twice as big  as
$H_{\Box}$. This implies  instability of the square VL with diagonals along
$a$ and $b$ (which is stable for $H_{\Box}<H<H_r$) with respect to
rotations. For fields $H>H_r$ we have four equilibrium square  VL's of the
same energy: the two with diagonals rotated relative to [100]
counter- and clockwise over an angle $\varphi$, and two for
rotations relative to [010]. We have  found by the direct numerical energy
minimization that $\varphi =0$ at $H=H_r$ and increases as $\sqrt{H-H_r}$
for $(H-H_r)\ll H_r$. For  LuNi$_2$B$_2$C   we
estimate that $\varphi$ may reach   $7\div 8$ degrees.

 \begin{figure}
\epsfxsize= 1.\hsize
\centerline{
\vbox{
\epsffile{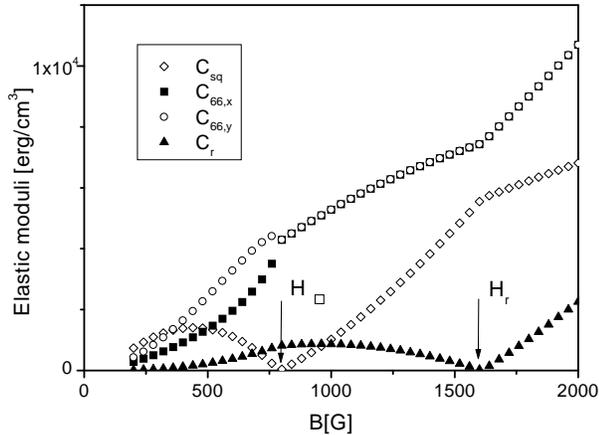}
}}
\vskip \baselineskip
\caption{The shear, rotation, and squash moduli versus field.}
\label{fig1}
\end{figure}

 \begin{figure}
\epsfxsize= .8\hsize
\centerline{
\vbox{
\epsffile{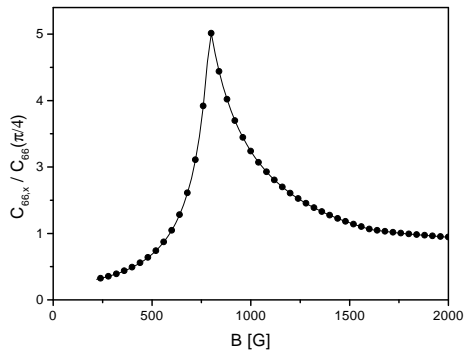}
}}
\vskip \baselineskip
\caption{The ratio $C_{66,x}/C_{66}(\pi/4)$ versus field.}
\label{fig2}
\end{figure}

  As Fig.1 shows, $C_{66,x}$ increases with field. For
$H<H_{\Box}$, $C_{66,x}\ne C_{66,y}$, whereas in the square phase between
$H_{\Box}$ and $H_r$ they  coincide, as they should by symmetry (the moduli
for $B\simeq 200\,$G are not plotted because the VL in domain differs from
that for $200\,$G$<B<H_{\Box}$). However, response to the shear other than of
$x$ or $y$ polarization,   differs from either $C_{66,x}$ or $C_{66,y}$. The
difference is maximum for the shear along [110] in the vicinity of the
transition field $H_{\Box}$, see Fig. 2.    \\
 
We thus come to a striking conclusion: weak nonlocal corrections to the 
London description suffice to make elastic properties of VL's strongly 
anisotropic even in the $ab$ plane where cubic or tetragonal materials are
macroscopically isotropic.  

 Softening of the squash mode near $H_{\Box}$ should affect
vortex fluctuations. This, however, does not lead to either
fluctuations divergence or to a peak in the critical
current\cite{Baruch}, the question recently discussed in Ref.
\onlinecite{Gurevich-Kogan}.  

 The rotation modulus $C_r$ vanishing at $H=H_r$ affects the equilibrium
VL structure above  $H_r$. Although some indications for the square VL
being unstable relative to rotations in sufficiently high fields were seen
in decorations of $\rm LuNi_2B_2C$ \cite{vinnik_private}, experimental
confirmation of this instability is still to be established. 

  The square VL is soft with respect to shear displacement along the
square sides (along [110] or $[{\bar 1}10]$). This suggests that the
vortex rows parallel to one of these directions can easily slide  by 
each other. If this happens for, e.g., [110] direction, the VL will possess
a long range order along [110], but will look disordered if observed
along $[{\bar 1}10]$. Such situations have been seen in decorations in
fields near $H_{\Box}$\cite{vinnik_private}.  The anisotropy
of the shear moduli can, in principle, be probed experimentally if the
critical current is arranged to flow only along [100] or [110] directions. 

Ames Laboratory is operated for US DOE by the Iowa State University under
Contract No. W-7405-Eng-82.

\end{multicols}
\end{document}